\documentclass[aps,prl,twocolumn,groupedaddress,showpacs,preprintnumbers
]{revtex4}
\usepackage{graphicx}
\usepackage{bm}

\begin{document}

\title{Viscoelasticity in normal $^{3}$He as a consequence of the 
Landau theory of normal Fermi liquids}

\author{Isadore Rudnick}
\altaffiliation{Deceased}
\author{Joseph Rudnick}
\affiliation{Department of Physics and Astronomy
University of California Los Angeles
Los Angeles CA 90095}

\date{\today}

\begin{abstract}

We show that a viscoelastic dispersion relation for shear waves in 
normal $^{3}$He at low temperatures follows directly from the Landau 
theory of normal Fermi liquids. This theoretical result is in accord 
with the experimental observations of propagating high frequency 
shear waves that identify normal $^{3}$He as a viscoelastic substance.
\end{abstract}


\maketitle

A viscoelastic substance has high frequency properties that are
characteristic of a solid an at low frequency the properties of a
viscous fluid.  The best known example is Silly Putty$^{\copyright}$. 
At very low temperatures, $^{3}$He in its normal state also has these
properties \cite{rudnick}.  Somewhat surprisingly it was only
recognized to be a viscoelastic substance several years after the
development of the Landau theory of normal Fermi systems
\cite{landau}.  $^{3}$He is also a remarkable viscoelastic liquid. 
Ordinary liquids exhibit viscoelasticity that is far from simple. 
Both shear and bulk moduli relax, and a spectrum of relaxation times
is commonly observed \cite{herzfeld}.  A viscoelastic substance in
which only the shear modulus relaxes, with a single relaxation time,
is called an ideal viscoelastic substance.  Based on experiments with
longitudinal sound, $^{3}$He appears to be an ideal viscoelastic
substance, and is, furthermore, the \emph{only known example} of such
a substance.  It has been shown on theoretical grounds \cite{bedell}
that this is true to a ``reasonable first approximation.''  However,
it is also found in in \cite{bedell} that ideal viscoelastic behavior
is ``only a crude first approximation for \emph{transverse} sound,''
and that the ideal ``viscoelastic model does not give quantitatively
reliable results.''  Special importance is attached to the fact that
the transverse sound velocity is less than or equal to the Fermi
velocity, and it is inferred that this interferes with the validity of
the viscoelastic interpretation of transverse zero sound.

The purpose of this note is to show that Fermi liquid theory can be 
used to derive the general viscoelastic equations for transverse zero 
sound exactly, independently of the relative magnitudes of the 
transverse sound velocity and the Fermi velocity. Thus, while $^{3}$He 
may not be an ideal viscoelastic substance for transverse sound, 
nevertheless it is a viscoelastic substance.

What we are able to do is show that the Landau theory leads to a 
dispersion relation for transverse zero sound that is \emph{exactly} 
as given by a viscoelastic model. This dispersion relation is
\begin{equation}
\left(\frac{\omega}{q} \right)^{2} = \frac{\mu_{\perp}(q,\omega)}{\rho}
\label{eq:1}
\end{equation}
where $\omega$ and $q$ are the frequency and wave vector of the 
transverse mode, $\rho$ is the mass density and $\mu_{\perp}(q, 
\omega)$ is the transverse modulus, i. e.
\begin{equation}
\Pi_{xz} = \mu_{\perp} \left( \frac{\partial u_{x}}{\partial z} + 
\frac{\partial u_{z}}{\partial x}\right)
\label{eq:2}
\end{equation}
where $\Pi_{xz}$ is an off-diagonal element of the stress tensor 
and $(\partial u_{x}/\partial z + \partial u_{z}/ \partial x)$ is the 
corresponding elemetnt of the strain tensor. The dependence on $q$ 
of $\mu_{\perp}$ arises from its non-locality in real space.

In the remainder of this note, we obtain from the Landau theory of
normal Fermi liquids a general expression for $\mu_{\perp}(q,\omega)$. 
We then derive a closed-form dispersion relation for propagating 
transverse modes, valid for all frequencies and relaxation times. 
Finally, we show that this dispersion relation can be cast in the 
form (\ref{eq:1}). This establishes the viscoelastic model of 
transverse zero sound propagation as a rigorous outcome of Landau 
theory. We note at the outset that two assumptions are made in 
obtaining our result. The first is that all Landau parameters $F_{l}$ 
with $l \ge 3$ are zero. The second is that all distortions of the 
Fermi surface decay locally via collisions with a single relaxation 
time, with the exception of a dilation and a translation, which do not 
decay via collisions because of particle and momentum conservation. 

Small deviations from equilibrium of the quasiparticle distribution
function, $n_{\vec{p}}(\vec{x},t)$ will be denoted by $\delta
n_{\vec{p}}(\vec{x},t)$, where $\vec{p}$ is the quasiparticle momentum
\cite{note}.  We assume that the mode has a space and time dependence
of the form $e^{i(qz-\omega t)}$.  The linearized Landau kinetic
equation is, then, in the absence of external forcing \cite{bedell},
\begin{equation}
\left( - i \omega + i qv_{F} \cos \theta \right) n_{\vec{p}} + 
\frac{\partial n_{\vec{p}}^{0}}{\partial \epsilon_{F}} \cos \theta 
\sum_{\vec{p}^{ \prime}}f^{s}_{\vec{p}\vec{p}^{\prime}} \delta 
n_{\vec{p}^{\prime}} = I \left[ n_{\vec{p}}\right]
\label{eq:3}
\end{equation}
The quantity $f^{s}_{\vec{p}\vec{p}^{\prime}}$ is the spin-symmetric 
quasiparticle interaction energy, $I[n_{\vec{p}}]$ is the collision 
contribution to the kinetic equation and $n_{\vec{p}}^{0}$ is the 
equilibrium distribution. We are interested in transverse modes, so we 
write
\begin{eqnarray}
\delta n_{\vec{p}} & = & - \frac{\partial n_{\vec{p}}^{0}}{\partial 
\epsilon_{\vec{p}}}\nu_{\vec{p}} \nonumber \\
& = & \frac{N(0)}{p_{F}^{2}} \delta (p-p_{F}) \nu_{\vec{p}}
\label{eq:4}
\end{eqnarray}
where $p_{F}$ is the Fermi momentum, 
\begin{equation}
\nu_{\vec{p}} = B \sin \theta \cos \phi + D \sin \theta \cos \theta 
\cos \phi + \bar{\nu}_{\vec{p}}
\label{eq:5}
\end{equation}
and $\bar{\nu}_{\vec{p}}$ consists of the contributions of all higher 
order, appropriately transverse, spherical harmonics to 
$\nu_{\vec{p}}$. Inserting (\ref{eq:4}) and (\ref{eq:5}) into 
(\ref{eq:3}), and adopting a single-relaxation-time model for 
$I[n_{\vec{p}}]$ we obtain
\begin{eqnarray}
\left[ - i \omega + i q v_{F} \cos \theta 
\left(1+\frac{F_{1}}{3}\right) \right] B \sin \theta \cos \phi 
\nonumber \\
+ \left[ - i \omega + i q v_{F} \cos \theta \left( 1+ 
\frac{F_{2}}{5}\right) + \frac{1}{\tau} \right] D \sin \theta \cos 
\theta \cos \phi \nonumber \\
+ \left[ - i \omega + i q v_{F} \cos \theta + \frac{1}{\tau} 
\right]\bar{\nu}_{\vec{p}}¥ & = & 
0 \nonumber \\
\label{eq:6}
\end{eqnarray}
We are assuming that $F_{l} =0$ when $l \ge 3$. The $F_{l}$'s are 
Landau parameters defined in the standard way. 

The modulus $\mu_{\perp}$ can now be obtained from (\ref{eq:3}) using
\begin{eqnarray}
\Pi_{xz} & = & \sum_{\vec{p}} p_{x}(v_{\vec{p}})_{z} \delta 
\bar{n}_{\vec{p}} \nonumber \\
& = & p_{F}v_{F} N(0) D \left(1+\frac{F_{2}}{5}\right) \int \sin^{3} 
\theta \cos ^{2} \theta \cos^{2} \phi \ d \theta \  d \phi \nonumber \\
& = & p_{F} v_{F} N(0) D \left( 1 + \frac{F_{2}}{5}\right) I_{1}
\label{eq:7}
\end{eqnarray}
where the last equality should be taken as a definition of $I_{1}$. 
In (\ref{eq:7}), we used
\begin{equation}
\delta \bar{n}_{\vec{p}} = \delta n_{\vec{p}} - \frac{\partial 
n_{\vec{p}}^{0}}{ \partial \epsilon_{\vec{p}}} 
\sum_{\vec{p}^{\prime}} f^{s}_{\vec{p}\vec{p}^{\prime}} \delta 
n_{\vec{p}^{\prime}}
\label{eq:8}
\end{equation}

The strain tensor, $\vec{u}_{xz}$ is calculated using the connection
between current density and strain \cite{bedell}.  One obtains
\begin{eqnarray}
\frac{\partial u_{x}}{\partial z} &  = & \frac{qN(0)}{\omega n}v_{F}B 
\left( 1+\frac{F_{1}}{3}\right) \int \sin^{3} \theta \cos ^{2} \phi \
d \theta \ d \phi \nonumber \\
& = & \frac{qN(0)}{\omega n}v_{F}B \left(1+ \frac{F_{1}}{3}\right) 
I_{2}
\label{eq:9}
\end{eqnarray}
The last equality above defines $I_{2}$. Thus,
\begin{eqnarray}
\mu_{\perp} & = & - \frac{\Pi_{xz}}{\partial u_{x}/\partial z} 
\nonumber \\
& = & - \frac{\omega n p_{F}}{q} \frac{D}{B} \frac{ 
1+\frac{F_{2}}{5}}{1+\frac{F_{1}}{3}} \frac{I_{1}}{I_{2}}
\label{eq:10}
\end{eqnarray}
The quantity $n$ in (\ref{eq:9}) and (\ref{eq:10}) is the particle 
number density.

The ratio $D/B$ in (\ref{eq:10}) is obtained by noting that 
$\bar{\nu}_{\vec{p}}$, which, according to (\ref{eq:6}) is given by
\begin{eqnarray}
\lefteqn{ \bar{\nu}_{\vec{p}} } \nonumber \\
& = & -\frac{\omega - qv_{F} \cos \theta \left(
1+\frac{F_{1}}{3}\right)}{\omega - q v_{F} \cos \theta +
\frac{i}{\tau}} B \sin \theta \cos \phi \nonumber \\
& & - \frac{\omega - qv_{F} \cos \theta \left( 1+ 
\frac{F_{2}}{5}\right) + \frac{i}{\tau}}{\omega - q v_{F} \cos 
\theta + \frac{i}{\tau}} D \cos \theta \sin \theta \cos \phi 
\nonumber \\
\label{eq:11}
\end{eqnarray}
is orthogonal to the spherical harmonic $Y^{1}_{2} \propto \cos 
\theta \sin \theta \cos \phi$. Thus,
\begin{equation}
\frac{D}{B} = \frac{- \int \frac{\left( \omega - qv_{F} \cos \theta 
\left( 1+\frac{F_{1}}{3}\right) \right) \cos \theta \sin^{3} \theta 
\cos^{2} \phi}{ \omega - qv_{F} \cos \theta + \frac{i}{\tau}} d \theta 
\ d \phi}{\int \frac{\omega - q v_{F} \cos \theta \left( 
1+\frac{F_{2}}{5} \right) + \frac{i}{\tau}}{\omega - qv_{F} \cos 
\theta + \frac{i}{\tau}} \cos^{2} \theta \sin^{3} \theta \cos^{2} \phi 
d \theta \ d \phi}
\label{eq:12}
\end{equation}
The modulus described by (\ref{eq:10}) with $D/B$ given by 
(\ref{eq:12}) interpolates between a low frequency limit proportional 
to $\omega \tau/i$ and a finite high frequency limit that is real and 
positive if $\omega > qv_{F}$ and complex otherwise. It is thus a 
\underline{viscoelastic} modulus.

Equation (\ref{eq:11}) also provides us with the dispersion relation 
for a propagating transverse mode. Notice that $\bar{\nu}_{\vec{p}}$ 
is also orthogonal to $Y_{1}^{1} \propto \sin \theta \cos \phi$. We 
then obtain the following requirement for a nontrivial solution to the 
undriven kinetic equation:
\begin{widetext}
\begin{eqnarray}
\lefteqn{\int  \frac{\omega - qv_{F} \cos \theta \left( 1+ 
\frac{F_{1}}{3}\right)}{\omega - qv_{F} \cos \theta + \frac{i}{\tau}} 
\sin^{3} \theta \cos ^{2} \phi \ d \theta \ d \phi} \nonumber \\
& = & \frac{\int \frac{\left( \omega - qv_{F}\cos \theta \left( 
1+\frac{F_{1}}{3}\right) \right) \cos \theta \sin^{3} \theta \cos^{2} 
\phi \ d \theta \ d \phi}{q-v_{F} \cos \theta + \frac{i}{\tau}} \int 
\frac{ \left( \omega - qv_{F} \cos \theta \left( 
1+\frac{F_{2}}{5}\right) + \frac{i}{\tau}\right) \cos \theta \sin^{3} 
\theta \cos^{2} \phi \ d \theta d \phi}{\omega -qv_{F} \cos \theta + 
\frac{i}{\tau}}}{\int \frac{\omega - qv_{F} \cos \theta \left( 
1+\frac{F_{2}}{5}\right) + \frac{i}{\tau}}{q-v_{F} \cos \theta + 
\frac{i}{\tau}} \cos^{2} \theta \sin^{3} \theta \cos^{2} \phi \ d 
\theta \ d \phi}
\label{eq:13}
\end{eqnarray}
\end{widetext}
This is a closed-form dispersion relation for undriven transverse 
modes in a normal Fermi liquid. Its solution interpolates between an 
overdamped low frequency mode and a propagating high frequency shear 
wave. 

A few further manipulations suffice to reduce this dispersion relation 
to the desired form (\ref{eq:1}). First, for the left hand side we have
\begin{eqnarray}
\lefteqn{\int \left( \frac{1}{\omega - q v_{F} \cos \theta + 
\frac{i}{\tau}} - \frac{1}{\omega + \frac{i}{\tau}}\right) \left( 
\omega - qv_{F} \cos \theta \left(1+\frac{F_{1}}{3}\right) \right) } 
\nonumber \\ &&
\sin^{3} \theta \cos^{2} \phi \ d \theta \ d \phi \nonumber \\
&&+ \frac{\omega}{\omega + \frac{i}{\tau}} \int \sin^{3} \theta 
\cos^{2} \phi \ d \phi \ d \theta \nonumber \\
& = &  \frac{qv_{F}}{\omega + \frac{i}{\tau}} \int \frac{\omega - qv_{F} 
\cos \theta \left(1+\frac{F_{1}}{3}\right)}{\omega - qv_{F} \cos 
\theta + \frac{i}{\tau}} \cos \theta \sin^{3} \theta \cos ^{2} \phi \ 
d \theta  \ d \phi \nonumber \\
& & + \frac{\omega}{\omega + \frac{i}{\tau}} I_{2}
\label{eq:14}
\end{eqnarray}
Inserting this into (\ref{eq:13}) we obtain, after some manipulations,
\begin{widetext}
\begin{eqnarray}
\omega I_{2} & = & \left\{ \int \frac{\left(\omega +
\frac{i}{\tau}\right) \cos \theta \sin^{3} \theta \cos^{2} \phi \left(
\omega -qv_{F} \cos \theta \left( 1+\frac{F_{2}}{5}\right) +
\frac{i}{\tau}\right) d \theta \ d \phi}{\omega - qv_{F} \cos \theta +
\frac{i}{\tau}} \right.  \nonumber \\ && \left.  - qv_{F} \int
\frac{\cos^{2} \theta \sin^{3} \theta \cos^{2} \phi \left( \omega -
qv_{F} \cos \theta \left(1+\frac{F_{2}}{5}\right) +\frac{i}{\tau}
\right) d \theta \ d \phi}{\omega - qv_{F} \cos \theta +
\frac{i}{\tau}} \right\} \nonumber \\
&& \times \left\{ \frac{ \int \frac{\omega - qv_{F} \cos \theta 
\left( 1+\frac{F_{1}}{3}\right)}{\omega -qv_{F} \cos \theta 
+\frac{i}{\tau}} \cos \theta \sin^{3}\theta \cos^{2} \phi \ d 
\theta \ d \phi}{\int \frac{\omega -qv_{F} \cos \theta \left( 
1+\frac{F_{2}}{5}\right) + \frac{i}{\tau}}{\omega - qv_{F} \cos 
\theta + \frac{i}{\tau}} \cos^{2} \theta \sin^{3} \theta \cos^{2} 
\phi \ d \theta \ d \phi} \right\}
\label{eq:15}
\end{eqnarray}
\end{widetext}
A few straightforward steps and reference to (\ref{eq:9}) and 
(\ref{eq:12}) yields
\begin{equation}
\omega = qv_{F}\left(1+\frac{F_{2}}{5}\right) 
\frac{I_{1}}{I_{2}}\frac{D}{B}
\label{eq:16}
\end{equation}
Thus,
\begin{equation}
\left(\frac{\omega}{q}\right)^{2} = \frac{\omega 
v_{F}}{q}\left(1+\frac{F_{2}}{5}\right) \frac{I_{1}}{I_{2}}\frac{D}{B}
\label{eq:17}
\end{equation}
Using (\ref{eq:10}):
\begin{equation}
\left(\frac{\omega}{q}\right)^{2} = \frac{1}{n} \frac{v_{F}}{p_{F}} 
\left(1+\frac{F_{2}}{5}\right) \mu_{\perp}(q,\omega)
\label{eq:18}
\end{equation}
Finally, since $1+F_{1}/3 = m^{*}/m$ and $v_{F}=p_{F}/m^{*}$, we have
\begin{eqnarray}
\left(\frac{\omega}{q}\right)^{2} & = & \frac{1}{mn}\mu_{\perp} 
\nonumber \\
& = & \frac{1}{\rho}\mu_{\perp}
\label{eq:19}
\end{eqnarray}
and the viscoelastic dispersion relation is recovered.

The above result establishes the viscoelastic model for transverse 
zero sound propagation in normal $^{3}$He as an exact consequence, 
within the assumptions noted above, of the Landau theory of normal 
Fermi liquids. However, we would like to point out that a propagating 
shear mode in liquid $^{3}$He is, in and of itself, evidence that the 
substance is viscoelastic. Such a mode occurs in solids and not in 
liquids. The criticisms base on the Landau theory of normal Fermi 
liquids have been of the modeling of transverse zero sound as an 
\emph{ideal} viscoelastic phenomenon. Because the Fermi velocity is 
greater than the velocity of transverse zero sound we do not expect 
the spatially local modulus with a single relaxation time that 
characerizes an ideal viscoelastic substance to apply here. As pointed 
out earlier, \emph{no} viscoelastic substance has been found to be 
ideal \underline{with the single exception of $^{3}$He} as a medium 
for longitudinal mode propagation. That the dispersion relation for 
transverse modes in $^{3}$He does not fit the ideal viscoelastic model 
is therefore not surprising. In our opinion, viscoelasticity provides 
an intuitively useful and quantitatively valuable model for zero 
sound propagation in normal $^{3}$He. Furthermore, as we have just 
seen, it follows directly from Landau theory.

\bibliography{visco}

\begin{thebibliography}{5}
\bibitem[{not()}]{note}
\bibinfo{note}{In the following our notation follows that of \cite{bedell}.
  Several results contained therein are also utilized.}
\end{thebibliography}

\end{document}